\documentclass[reprint,prl,amssymb,amsmath,aps,floatfix]{revtex4-1}
\usepackage{amsmath,amsthm,amssymb,amsfonts}
\usepackage{bm}%
\usepackage{graphicx,color}
\usepackage{hyperref}%
\usepackage{xfrac}

\usepackage[normalem]{ulem}
\expandafter\ifx\csname package@font\endcsname\relax\else
 \expandafter\expandafter
 \expandafter\usepackage
 \expandafter\expandafter
 \expandafter{\csname package@font\endcsname}%
\fi
\begin{document}
\title{Diffusion of active particles in a complex environment: role of surface scattering}
\author{Theresa Jakuszeit}
\email{tj295@cam.ac.uk}
\author{Ottavio A. Croze}
\author{Samuel Bell}
\email{sb855@cam.ac.uk}
\affiliation{Cavendish Laboratory, University of Cambridge, Cambridge CB3 0HE, U.K.}
\date{\today}
\begin{abstract}
Experiments have shown that self-propelled particles can slide along the surface of a circular obstacle without becoming trapped over long times. 
Using simulations and theory, we study the impact of boundary conditions on the diffusive transport of active particles in an obstacle lattice. 
We find that particle dynamics with sliding boundary conditions result in large diffusivities even at high obstacle density, unlike classical specular reflection. 
These dynamics are very well described by a model based on Run-and-Tumble particles with microscopically derived reorientation functions arising from obstacle-induced tumbles. This model, however, fails to describe fine structure in the diffusivity at high obstacle density predicted by simulations. Using a simple deterministic model, we show that this structure results from particles being guided by the lattice. Our results thus show how non-classical surface scattering introduces a dependence on the lattice geometry at high densities. We discuss implications for the study of bacteria in complex environments.
\end{abstract}

\flushbottom
\maketitle

The field of active matter covers the broad spectrum of particles which move by consuming energy from their environment~\cite{Bechinger2016}. These range from flocks of birds and insect swarms~\cite{Puckett2015,Sinhuber2017}, to cell tissues~\cite{Giavazzi2018}, microswimmers~\cite{Cates2012}, microtubuli~\cite{Sanchez2012,Keber2014}, and enzymes~\cite{Jee2018}. Microswimmers such as bacteria and Janus particles self-propel at low Reynolds numbers, the latter being directly powered by an asymmetric chemical reaction on the particle surface, the former by rotating helical filaments. The propulsive mechanisms set up complicated hydrodynamic flows, which determine the characteristics of interactions, both with other microswimmers, and with the boundaries of their environment. These boundary interactions may perform an essential function in nature. Surface-induced accumulation is an important step in the formation of biofilms, which are involved in many chronic diseases and pathogen spread~\cite{Costerton1999, Drescher2011}. Blood pathogens are adapted to swimming in crowded environments~\cite{Heddergott2012}, sperm cells follow the wall of the genital tract to reach the egg cell~\cite{Eisenbach2006,Guidobaldi2014,Denissenko2012}, and artificial Janus particles have been guided along microfluidic edges~\cite{Simmchen2016} and through obstacle arrays~\cite{Volpe2011,Brown2016, Wykes2017}.

The nature of particle-surface interactions relies on a microswimmer's propulsion mechanism, including steric and hydrodynamic effects. Microalgae, which are ``puller'' type swimmers, are scattered off surfaces~\cite{Kantsler2013ciliary,Contino2015,Lushi2017}, leading to billiard-like motion in polygon structures~\cite{Spagnolie2017Billiard}. In contrast, ``pusher'' type swimmers, such as bacteria or Janus particles, are trapped by hydrodynamic effects near flat surfaces, where they accumulate~\cite{Berke2008,Li2009accumulation,Elgeti2010}. When the surface is instead convex, this trapping time can be reduced~\cite{Sipos2015}. In particular, bacteria trace along convex surfaces such as microfluidic pillars before escaping with a small angle~\citep{Spagnolie2015}.

The modelling of these scenarios typically follows one of two approaches: 
hydrodynamic models, or 
random walk models. With a full hydrodynamic approach, the particle-surface interactions can be studied by modelling the active particle as a hard sphere with defined tangential surface velocity~\cite{Elgeti2015}. 
A recent study explored the migration of active particles through a body-centered cubic lattice of spheres of the same size as the particle~\cite{Chamolly2017}. Depending on the swimmer type and packing density, the authors found trapped, random walk and straight trajectories. The computational demands of the simulations, however, prevented study of long-time behavior. Random walk models can be used to study the diffusive behavior of active particles. Diffusion in complex media has been studied for several boundary interactions: for model particles that evade obstacles~\citep{Chepizhko2013}, particles that are trapped before being randomly reorientated~\cite{Bertrand2017}, and particles that interact with obstacles via an excluded volume potential~\cite{Zeitz2017}.
Hydrodynamic boundary interactions have been shown to play an important role in active systems, e.g. in the control of flow-induced phase separation \cite{Thutupalli2018}. Similarly, pusher-type boundary interactions may guide microswimmers through their environment~\cite{Simmchen2016, Raatz2015}, which would facilitate diffusion.

In this Rapid Communication, we study theoretically how the diffusive transport of active particles in ordered arrays of obstacles is influenced by boundary scattering. We consider particles specularly reflected from boundaries, as in the Lorentz gas model \cite{Machta1983}, and particles that scatter by sliding around obstacles, like pushers~\cite{Sipos2015, Spagnolie2015}. For these `pusher-like' collisions, our simulations and a run-and-tumble particle model we develop predict, counterintuitively, that large diffusive transport is possible even at high obstacle densities. This result contrasts sharply with the expected low diffusivity of Lorentz gas particles at high densities. We show, using a simple deterministic model, how this large diffusion at high density is caused by particle guiding by the lattice. Our results highlight the previously unexplored role of lattice geometry in active particle transport.

\textit{Model.} We consider $N_P$ active particles in a two-dimensional space in which obstacles are placed in a hexagonal lattice. The centers of the obstacles are fixed with distance $d$, and the obstacle radius $R$ is varied. The equations of motion for the $i$-th particle are given by
\begin{align}
\dot{\mathbf{x}}_i &=v \, \mathbf{p}(\varphi_i) \label{ActiveModelPosition} \\
\dot{\varphi}_i &=\sqrt{2 D_R} \xi_i(t),
\label{eq: ActiveModelDirection}
\end{align}
where dot denotes the time derivative, $v$ is the particle speed, $\mathbf{x}_i$ and $\varphi_i$ correspond to the position and moving direction of the $i$-th particle, respectively, and the unit vector $\mathbf{p}=[\cos\varphi,\sin\varphi]$. The white noise in Eq. \eqref{eq: ActiveModelDirection} obeys $\langle \xi (t) \rangle =0$ and $\langle \xi_i(t) \xi_j (t') \rangle= \delta_{ij} \delta (t-t')$. Thus, the moving direction undergoes rotational diffusion with $\langle \varphi(t)^2 \rangle=2D_R t$.
As a result, the particle performs a persistent random walk with persistence length $l_p=v/D_R$ \cite{Berg1993}. 

\begin{figure}
	\centering
	\includegraphics[width=\columnwidth]{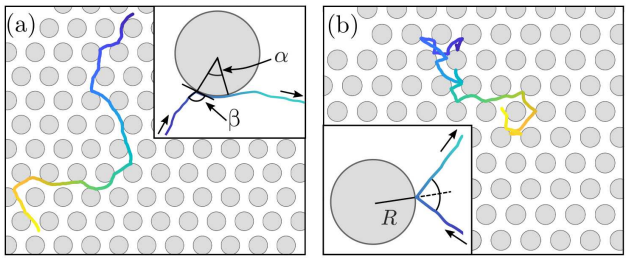}
	\caption{Boundary conditions. Typical trajectory of a particle governed by Eqs. \eqref{ActiveModelPosition} and \eqref{eq: ActiveModelDirection} with (a) a sliding, or
	(b) a reflecting boundary condition. 
	See text for details.}
	\label{fig:illustration}
\end{figure}

To capture the non-classical particle-surface interaction, we introduce a sliding boundary condition as illustrated in Fig. \ref{fig:illustration}(a) inset \cite{Sipos2015}. Consider a collision with an obstacle: $\beta$ is defined as the angle between the tangent at the collision point and the orientation $\mathbf{p}$. If $\beta < \pi/2$, the particle travels clockwise around the obstacle; if $\beta \geq \pi/2$, the particle travels counter-clockwise. The particle moves along the obstacle to traverse a central angle $\alpha$. Recent microfluidic experiments \cite{Sipos2015} and hydrodynamic models \cite{Sipos2015, Spagnolie2015} have shown that pillars with radii above a critical threshold strongly trap pushers, which escape at long times by rotational diffusion. We consider in this study only pillars with radii below this critical threshold. In this case, swimmers collide with an obstacle with angle $\beta$ defined above, slide along the surface and leave it after traversing a central angle $\alpha$ (Fig. \ref{fig:illustration}(a) inset). As the swimmer slides, the angle between it and the obstacle surface tangent decreases until escape~\cite{Spagnolie2015}. A model of stochastic dynamics could determine, for a given incident $\beta$, the resulting distribution of central angles $\alpha$ (leaving times). However, such a model has yet to be developed. In this work we thus explore the effect of boundary conditions assuming a fixed central angle $\alpha$ and further assume that, when a particle leaves an obstacle, its orientation $\mathbf{p}$ is tangent to the obstacle surface. This is a necessary simplification of the behaviour of pusher-type particles at convex obstacles. The neglect of stochasticity in $\alpha$ can be checked by simulations. Results (not shown) with a fixed (mean) $\alpha$ are qualitatively the same to those obtained with a distribution of $\alpha$, provided the latter is peaked about its mean (e.g. a Gamma distribution).

As a comparison, we also consider a reflecting boundary condition, where a particle is reflected with an angle equal to the incident angle, as illustrated in Fig.~\ref{fig:illustration}(b). This interaction type implies time-reversability, which is an assumption underlying gas kinetic models derived for bacteria transport in porous media \cite{Barton1997,Ford2007}. By contrast, the sliding boundary condition is not time reversible and violates detailed balance~\cite{Cates2012}. The system of Eqs. \eqref{ActiveModelPosition} and \eqref{eq: ActiveModelDirection} is solved numerically, and example particle tracks are shown in Fig.~\ref{fig:illustration}. We derive the diffusion coefficient from $N_P$ simulated particle tracks by fitting the mean square displacement as $\langle \delta x(t)^2\rangle = 4 D_\mathrm{eff} t +4D_\mathrm{eff} \eta [\exp (-t/\eta)-1]$ (a result easily derived for self-propelled particles using a standard method, see for example~\cite{Risken1984}), where the time scale of ballistic motion, $\eta$, is the second fitting parameter.

\begin{figure*}
	\centering
	\includegraphics[width=\textwidth]{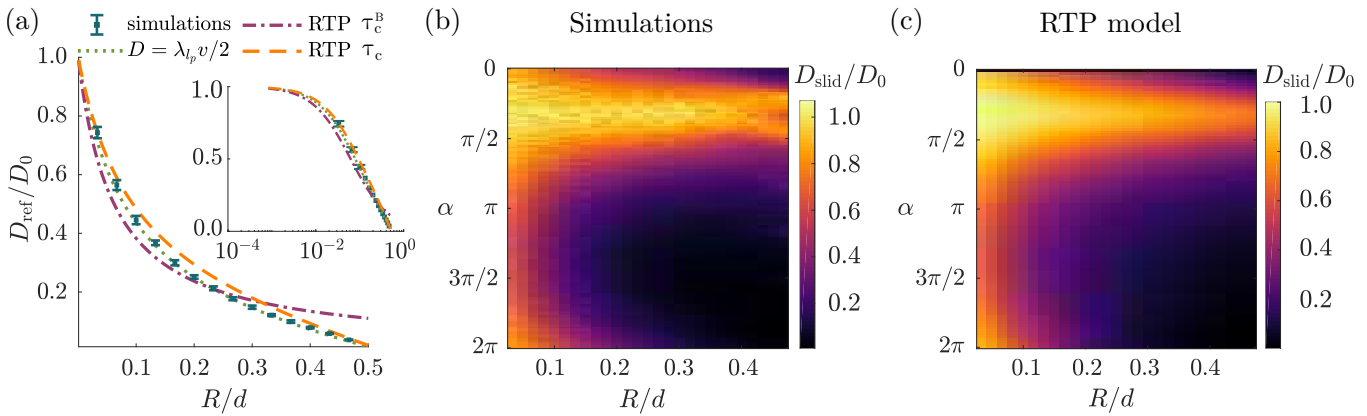}
	\caption{
	Diffusion in hexagonal lattice of obstacles.  (a) Diffusion with reflective boundary condition, $D_\mathrm{ref}$, is scaled by diffusion coefficient in the absence of any obstacles, $D_0=v^2/2D_R$. Simulations agree with Santalo's formula that was adjusted for rotational diffusion, $\lambda_{l_P}$ (green dotted curve). The run-and-tumble model in Eq. \eqref{eq:DiffusionRT} with Santalo mean free path $\lambda$ in $\tau_c=\lambda/v$ (orange dashed) is compared to RTP model with $\tau_c^B=1/\rho$ \cite{Bertrand2017} (purple dashed-dotted) and $\langle \cos \psi \rangle=-1/3$. 
	(b) Simulations with sliding boundary condition reveal dependence on both obstacle density and central angle. 
	(c) Theoretical prediction Eq.~\eqref{eq:DiffusionRT} with $\tau_c=\lambda/v$ in $\tau=\tau_c + \tau_R$ and $\langle \cos\psi\rangle$ given by Eq.~\eqref{eq:CosTheta}. Parameters: $N_P=1000$, $D_R=0.1~\text{s}^{-1}$, $v=20~\mu$m$\, \text{s}^{-1}$, $d=60~\mu$m.}
	\label{fig:ReflectiveSlidingResults}
\end{figure*}

\textit{Reflecting boundary condition.} We first establish the diffusive properties of active particles with a reflective boundary condition. Here, we recognize an analogy to the Lorentz gas model, in which particles move ballistically between obstacles \cite{Machta1983}. The Santalo formula is a well-known result for the mean-free path of a Lorentz gas \cite{Chernov1997} given by $\lambda=\pi A /P$, where $A$ and $P$ are the free area and obstacle perimeter in a unit cell, respectively.
Since the active particles move diffusively at large time scales, we derive an active version of Santalo's formula with a circle of radius $l_p$ as an additional boundary. This yields the mean-free path of an active particle as $\lambda_{l_p}=2 \pi N A /(N P + 2 \pi l_p)$,  where $N$ is the number of unit cells included in the circle of radius $l_p$. For a hexagonal lattice of circular obstacles, we obtain $A=\sqrt{3}d^2/2-\pi R^2$, $P=2\pi R$ and $N=\pi l_p^2/(\sqrt{3}d^2/2)$ \cite{[{See Supplemental Material at }] [{ for supporting figures illustrating the active version of Santalo formula and RTP model.}]SuppMat}. As shown in Fig.~\ref{fig:ReflectiveSlidingResults}(a), applying this adjusted mean-free path in $D=\lambda_{l_p}v/2$ matches the simulations.
The inset plots the theoretical prediction and the diffusion coefficient fitted from simulations on a lin-log scale, showing that at large $R/d$
the diffusion coefficient scales as $\ln(1/\rho)$, where obstacle density $\rho=2\pi/\sqrt{3} \, (R/d)^2$. 

\textit{Sliding boundary condition.} By contrast, numerical solutions of Eqs. \eqref{ActiveModelPosition} and \eqref{eq: ActiveModelDirection} with a sliding boundary condition reveal that diffusion depends both on the obstacle density $\rho$ and the central angle $\alpha$ [see Fig. \ref{fig:ReflectiveSlidingResults}(b)]. Surprisingly, a large diffusive transport can be sustained even at large obstacle density $\rho$ for certain values of $\alpha$. 
Despite frequent obstacle collisions, the reorientation is small because the sliding boundary condition conserves the major component of the velocity vector for small to intermediate values of $\alpha$.
Large values of $\alpha$, on the other hand, cause a particle to retrace much of its track. 
The typical pusher surface interaction can, thus, lead to an increase in effective diffusion compared to the classical reflection.

\textit{Theoretical framework.} While the active Santalo formula matches the reflective simulations well in Fig.~\ref{fig:ReflectiveSlidingResults}(a), it cannot account for the persistence introduced by the sliding boundary condition, and a different approach is required.
 We derive a theoretical description based on the model of run-and-tumble particles (RTP) \cite{Tailleur2008, Schnitzer1993}. In this Rapid Communication, an effective `tumble' is defined as an obstacle-induced reorientation of the particle, and the `run' between obstacle collisions is influenced by rotational diffusion. The diffusion coefficient for an RTP also undergoing rotational diffusion is known to be
\begin{equation}
D=\dfrac{v^2} {2[D_R + (1-\langle \cos\psi \rangle)/\tau]},
\label{eq:DiffusionRT}
\end{equation} 
where $\tau$ is the mean run time and $\psi=\psi\bm{(}\alpha,P(\beta)\bm{)}$ is the reorientation angle during a tumble \cite{Taktikos2013, Lovely1975}. The reorientation angle is the combination of alignment upon collision with the obstacle, $\beta$, and sliding according to the central angle, $\alpha$: $\psi=\alpha-\beta$. The average $\langle\cos\psi\rangle$ is performed over the collision angle $\beta$, with probability distribution $P(\beta)$. To derive the distribution, we assume that a particle can start at any point in free space with uniform distribution of directions, and then travels in a straight line. The probability distribution of a collision angle $\beta$ at a given distance $x$, $P_x(\beta)$, can be written in terms of $P(\phi)$, where $\phi(\beta,x)$ is the angle between $x$ and the moving direction. Thus, $P_x(\beta)d\beta=P(\phi)d\phi$.
For circular obstacles, $\phi(\beta,x)$ follows geometrically from the sine rule so that $\phi(\beta,x)=\sin^{-1}\left(R \, \cos \beta/x\right)$ \cite{SuppMat}. Differentiation yields the Jacobian $\left\lvert d\phi/d\beta\right\rvert=\left\lvert d\beta/d\phi\right\rvert^{-1}=R\sin(\beta)(x\sqrt{1-R^2 \, \cos^{2}\beta/x^2})^{-1}.$
Finally, we average over all initial positions
\begin{equation}\label{eq:pbeta}
P(\beta) =\lim_{L\to\infty} \frac{\int_{R}^{L}\;2\pi x \left\lvert\frac{d\phi}{d\beta}\right\rvert dx}{\int_{R}^{L}dx\int_{0}^{\pi}\;2\pi x \left\lvert\frac{d\phi}{d\beta}\right\rvert d\beta}=\frac{\sin\beta}{2},
\end{equation}
where $L$ is the system size. Despite using deterministic trajectories to calculate this distribution, it fits the observed collision angle distribution for simulations at low densities.
Performing the average gives the reorientation function as:
\begin{align}
\langle\cos\psi\rangle&=2\int_{0}^{\pi/2} \cos(\alpha-\beta) P(\beta) d\beta \nonumber \\
&=\frac{1}{4}(2\, \cos\alpha+\pi\sin\alpha),
\label{eq:CosTheta}
\end{align}
noting that $\cos\psi$ is even about $\beta=\pi/2$. For the reflecting boundary condition, $\psi=2\beta$, and $\langle \cos\psi \rangle=-1/3$. 

The second parameter in the RTP model \eqref{eq:DiffusionRT} is the mean run time $\tau$, which corresponds to the time between obstacle collisions. Because the characteristic time between collisions is independent of the details of the random walk and depends purely on confinement \cite{Blanco2003}, we use the mean collision time $\tau_c=\lambda/v$, where $\lambda$ is the mean free path given by Santalo's formula.  
For the sliding boundary condition, the mean run time is adjusted by the time spent on an obstacle, i.e. $\tau=\tau_c + \tau_R$, with residence time $\tau_R=R \alpha/v$. Travelling on the obstacle causes an effective reduction in velocity. When the particle traces along the pillar, it travels a distance $l< v \tau_R$, which gives $v_\mathrm{obs}=l/\tau_R$. By the cosine rule, $l = R\sqrt{2-2\cos\alpha}.$
The effective speed in Eq. \eqref{eq:DiffusionRT} is then $v_\mathrm{eff}=v \tau_c/\tau + v_\mathrm{obs} (\tau-\tau_c)/\tau$. 

\begin{figure*}
	\centering
	\includegraphics[width=\textwidth]{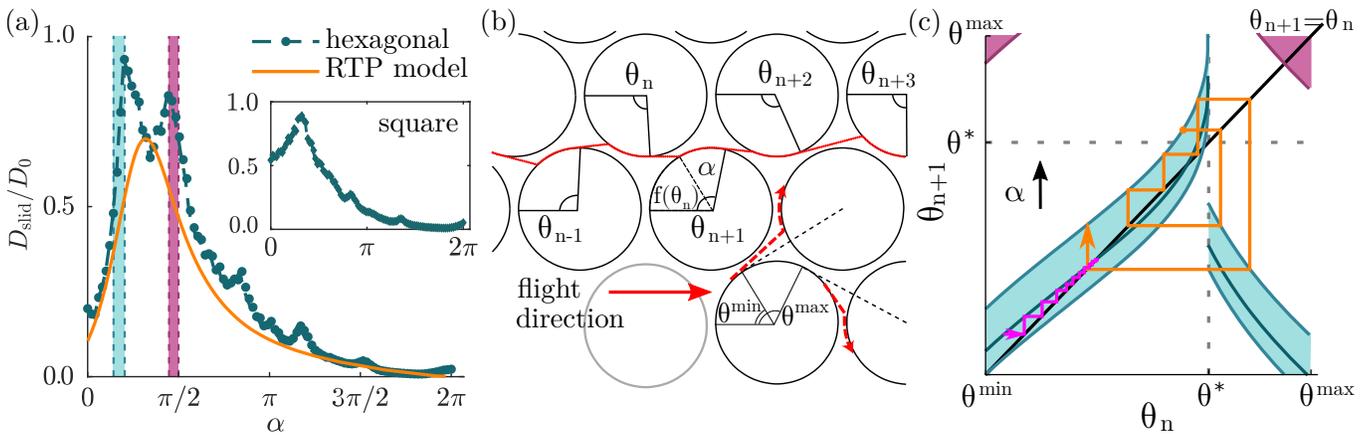}
	\caption{Geometric effects. (a) The discrepancy between the RTP model and the hexagonal lattice simulation results at high density  ($R/d=0.47$) is centred around the deterministic stable regions [shaded as in (c)], revealing influence of geometry. Inset: Diffusion coefficient for a square lattice. (b) Schematic of a 1-D system, considering a flight along one channel in the lattice. The leaving angle at each pillar is given by $\theta_n$. The lower schematic shows possible termination of  flights in a horizontal channel. 
	(c) Iterative map of the leaving angle as a function of the previous leaving angle for different central angles $\alpha$, $\theta_{n+1}=f(\theta_n)+\alpha$. The shaded regions correspond to regions of stable flights. Stable fixed points cross the dashed $\theta_{n+1}=\theta_n$ line with a gradient between -1 and 1 (a mapping with a stable fixed point is shown in the lower shaded region, with an example trajectory in pink). An example of a bounded mapping of leaving angles is shown as orange trajectory.}
	\label{fig:TheoryGeometry}
\end{figure*}

We first apply the RTP theory to simulations with reflecting boundary condition, using $\langle \cos\psi \rangle=-1/3$ and $\tau_R=0$. As shown in Fig. \ref{fig:ReflectiveSlidingResults}(a), the RTP model with $\tau=\tau_c$ yields a good approximation of the simulation results. As a comparison, the RTP model with a recently derived mean collision time \cite{Bertrand2017}, where $\tau_c^B=1/\rho$, approximates the simulations at low densities but diverges in the high density regime.

For the sliding boundary condition, the RTP framework reproduces the main features of the simulations, see Figs.~\ref{fig:ReflectiveSlidingResults}(b) and \ref{fig:ReflectiveSlidingResults}(c): it maintains a large diffusion coefficient for small to intermediate $\alpha$. Since $\tau_c$ is independent of the boundary condition, this must stem from the reorientation function $\langle \cos\psi \rangle$ in Eq.~\eqref{eq:CosTheta}, which has a maximum at $\alpha\approx \pi/3$ and a minimum at $\alpha \approx 4\pi/3$. These extrema coincide with the predicted maximum and minimum of the diffusion coefficient observed for small to intermediate $R/d$ in Fig.~\ref{fig:ReflectiveSlidingResults}(c). Beyond $\alpha=4\pi/3$, any increase in the diffusion coefficient due to the reorientation function is suppressed by the increase in residence time $\tau_R$ at large $R$ and $\alpha$.
Note that, since the RTP model is oblivious to obstacle arrangement, these results also apply to random lattices at low densities.

\textit{High density geometrical effects.} While the RTP model accounts for the diffusion coefficient $D_\mathrm{slid}$ at low to intermediate obstacle densities, it fails to completely describe the simulations at high density. Figure~\ref{fig:TheoryGeometry}(a) shows fixed $R/d=0.47$ (the largest value) cross-sections of the surfaces in Figures~\ref{fig:ReflectiveSlidingResults}(b) and~\ref{fig:ReflectiveSlidingResults}(c). At this high density, the diffusion coefficient for the hexagonal lattice simulations has peaks that exceed the RTP model. There are two of these peaks at low $\alpha$ as well as smaller overshoots at higher $\alpha$. However, if we instead perform the simulations in a square lattice, we get a different peak structure, with a single peak at low $\alpha$. We will show that this is due to the geometry of the lattice, and its guiding effect on the self-propelled particles.

For the geometry of the lattice to influence the particle paths, there must be a correlation between successive collisions  with pillars. This means that the particle must not loose the memory of its orientation between collisions, i.e. the obstacle separation must be much smaller than the persistence length, $d-2R\ll l_p$. In this case, a purely deterministic model ($D_R=0$) provides a good approximation to explore correlations between collisions. In such a model, the particle travels in a straight line between pillars, and is reoriented by $\alpha$ by sliding scattering. We consider a `channel' defined by two rows of pillars within the lattice (Fig.~\ref{fig:TheoryGeometry}(b)). A particle traverses the channel by skirting around pillars, leaving the surface of the $n$th pillar with a polar angle $\theta_n$. For deterministic (ballistic) dynamics between collisions, we can completely specify a trajectory by the `flight' $\lbrace\theta_n\rbrace_{n=1}^N$, the sequence of leaving angles from successive collisions, as in Fig.~\ref{fig:TheoryGeometry}(b). The sequence size $N$ defines the flight length. Successive leaving angles are determined by the recurrence relation: $\theta_{n+1}=g_\alpha(\theta_n)=f(\theta_n)+\alpha$, where, in this deterministic model, $f(\theta_n)$ is a function determined solely by the geometry.

As the particle moves along a channel during a flight, it can transition between pillars on the opposite ( e.g. $\theta_{n-1}\to \theta_n$) or same (e.g. $\theta_{n+2}\to \theta_{n+3}$) side of the channel, as shown Fig.~\ref{fig:TheoryGeometry}(a). For $R/d>\sqrt{3}/4$ (the close-packed limit of overlapping pillars), a critical angle $\theta_n=\theta^*$ emerges that determines on which side of the channel a particle will next hit. If $\theta_n\leq\theta^*$, the particle will cross over to an obstacle on the other side of the channel, while if $\theta_n>\theta^*$, it will move to one on the same side. This means the map $g_\alpha(\theta_n)$ is discontinuous at $\theta_n=\theta^*$, as in Fig.~\ref{fig:TheoryGeometry}(c).

The flights considered in the deterministic model correspond well to what we observe in our simulations. At high densities, these show particle trajectories made up of long flights along lattice channels, interrupted by `tumbles' into the next long flight. The deterministic model allows to establish if the flights are geometrical in origin. In this model, a flight terminates when the leaving angle $\theta_n$ becomes too small ($\theta_n<\theta^\mathrm{min}$) or too large ($\theta_n>\theta^\mathrm{max}$) as it will be deflected out of the channel on its next collision, illustrated in Fig.~\ref{fig:TheoryGeometry}(b). Stable flights are trajectories that remain in the region $\theta^\mathrm{min}\leq\theta_n\leq\theta^\mathrm{max}$ indefinitely. This can happen in two ways: (i) a stable fixed point may exist (a point $\theta$ such that $g_\alpha(\theta)=\theta$, and $|g'_\alpha(\theta)|<1$), so that long trajectories have a single repeated leaving angle; (ii) the map $g_\alpha(\theta_n)$ is bounded within the allowed region of leaving angles: $\theta^\mathrm{min}\leq g_\alpha(\theta_n)\leq\theta^\mathrm{max}$ for all $\theta^\mathrm{min}\leq\theta_n\leq\theta^\mathrm{max}$, so that no trajectory may leave the allowed region. Example trajectories of both types are illustrated in Fig.~\ref{fig:TheoryGeometry}(c).

The iterative map $\theta_{n+1}=g_\alpha(\theta_n)$ is plotted for $R/d=0.47$ in Fig.~\ref{fig:TheoryGeometry}(b). Two stable ranges (shaded regions) are seen to emerge corresponding to ranges of $\alpha$, which controls stability. For $\theta_n<\theta^*$, increasing $\alpha$ causes a stable fixed point to develop. Increasing it further, in the range that defines the lower region (shaded in blue), provides a map bounded in the interval $[\theta_{min}, \theta_{max}]$. Flights in this lower shaded region bounce from one side of the channel to the other. If $\alpha$ is increased further, the map again becomes unbounded ($g_\alpha(\theta_n)>\theta^\mathrm{max}$) and stability is lost. For $\theta_n>\theta^*$, the upper region (shaded in pink) has a stable fixed point, so that particles perform stable flights by running along only one side of the channel in this region.
Stable trajectories from the deterministic model cannot give rise to diffusive behaviour. However, any rotational diffusion, however small, will eventually cause a deviation of trajectory large enough to take the particle out of the stable interval $[\theta^\mathrm{min},\theta^\mathrm{max}]$. This will cause flights to terminate, giving rise to diffusive behaviour. In view of the large persistence length of flights for stable values of $\alpha$, the diffusion coefficient for such flights is expected to be large compared to that corresponding to other values of $\alpha$. By plotting the stable regions of $\alpha$ predicted by the deterministic model against the simulation results at high density in Fig.~\ref{fig:TheoryGeometry}(a), we see that this is indeed the case: the spikes in diffusion coefficient for the simulations correspond well to the stable regions in the deterministic model. It is important to note that the obstacle sizes we are considering here are below the critical trapping radii tyically found~\cite{Sipos2015,Spagnolie2015}. It is possible to reach a high density state where the obstacle separation is larger than the persistence length, where our results wouldn't hold. However, in this regime, the obstacles would be much larger than the trapping radius, and so particles would be trapped for long periods on obstacles~\cite{Sipos2015,Spagnolie2015}, making diffusion very slow. 

To conclude, we find that non-classical surface interactions significantly impact the active diffusive transport in complex environments, such as ordered obstacle arrays. Our results highlight the importance of choosing realistic microscopic boundary conditions to obtain realistic macroscopic dynamics. In particular, models employing reflective boundary conditions, e.g. those used in \cite{Barton1997,Ford2007} to describe bacteria in porous media, should not give realistic results for active particles. While this is generally obvious considering detailed balance~\cite{Cates2012}, the theoretical framework we have developed allows the formulation of particular predictions to be tested experimentally. e.g. using bacteria in microfluidic arrays. In particular, it would be interesting to test our prediction of large diffusive transport in dense arrays. While the description was developed for lattices, we note that, when the number of obstacle contacts is low, our results hold for random environments too.
 

Finally, since the time bacteria spend on an obstacle is a function of its curvature and the force dipole strength of the bacterium \citep{Sipos2015,Spagnolie2015}, it is interesting to consider the diffusive transport of bacterial species with different dipole strengths. The latter depend on body shape and propulsion mechanism, which vary between species. It would be interesting to investigate if certain species, e.g. soil bacteria, have hydrodynamic properties tailored towards guided transport in complex environments\cite{Raatz2015}. This could be achieved combining our theoretical framework and microfluidic experiments.

\begin{acknowledgments}
We thank Eugene Terentjev, Mark Warner and Mike Cates for helpful discussions and feedback on the manuscript. This work has been funded by EPSRC EP/M508007/1 (S.B.), EP/L504920/1 and EP/N509620/1 (T.J.), 
and the Winton Programme for the Physics of Sustainability (T.J., O.C.). 
\end{acknowledgments}

\bibliographystyle{apsrev4-1}

\begin{thebibliography}{46}%
\makeatletter
\providecommand \@ifxundefined [1]{%
 \@ifx{#1\undefined}
}%
\providecommand \@ifnum [1]{%
 \ifnum #1\expandafter \@firstoftwo
 \else \expandafter \@secondoftwo
 \fi
}%
\providecommand \@ifx [1]{%
 \ifx #1\expandafter \@firstoftwo
 \else \expandafter \@secondoftwo
 \fi
}%
\providecommand \natexlab [1]{#1}%
\providecommand \enquote  [1]{``#1''}%
\providecommand \bibnamefont  [1]{#1}%
\providecommand \bibfnamefont [1]{#1}%
\providecommand \citenamefont [1]{#1}%
\providecommand \href@noop [0]{\@secondoftwo}%
\providecommand \href [0]{\begingroup \@sanitize@url \@href}%
\providecommand \@href[1]{\@@startlink{#1}\@@href}%
\providecommand \@@href[1]{\endgroup#1\@@endlink}%
\providecommand \@sanitize@url [0]{\catcode `\\12\catcode `\$12\catcode
  `\&12\catcode `\#12\catcode `\^12\catcode `\_12\catcode `\%12\relax}%
\providecommand \@@startlink[1]{}%
\providecommand \@@endlink[0]{}%
\providecommand \url  [0]{\begingroup\@sanitize@url \@url }%
\providecommand \@url [1]{\endgroup\@href {#1}{\urlprefix }}%
\providecommand \urlprefix  [0]{URL }%
\providecommand \Eprint [0]{\href }%
\providecommand \doibase [0]{http://dx.doi.org/}%
\providecommand \selectlanguage [0]{\@gobble}%
\providecommand \bibinfo  [0]{\@secondoftwo}%
\providecommand \bibfield  [0]{\@secondoftwo}%
\providecommand \translation [1]{[#1]}%
\providecommand \BibitemOpen [0]{}%
\providecommand \bibitemStop [0]{}%
\providecommand \bibitemNoStop [0]{.\EOS\space}%
\providecommand \EOS [0]{\spacefactor3000\relax}%
\providecommand \BibitemShut  [1]{\csname bibitem#1\endcsname}%
\let\auto@bib@innerbib\@empty
\bibitem [{\citenamefont {Bechinger}\ \emph {et~al.}(2016)\citenamefont
  {Bechinger}, \citenamefont {Di~Leonardo}, \citenamefont {L{\"o}wen},
  \citenamefont {Reichhardt}, \citenamefont {Volpe},\ and\ \citenamefont
  {Volpe}}]{Bechinger2016}%
  \BibitemOpen
  \bibfield  {author} {\bibinfo {author} {\bibfnamefont {C.}~\bibnamefont
  {Bechinger}}, \bibinfo {author} {\bibfnamefont {R.}~\bibnamefont
  {Di~Leonardo}}, \bibinfo {author} {\bibfnamefont {H.}~\bibnamefont
  {L{\"o}wen}}, \bibinfo {author} {\bibfnamefont {C.}~\bibnamefont
  {Reichhardt}}, \bibinfo {author} {\bibfnamefont {G.}~\bibnamefont {Volpe}}, \
  and\ \bibinfo {author} {\bibfnamefont {G.}~\bibnamefont {Volpe}},\
  }\href@noop {} {\bibfield  {journal} {\bibinfo  {journal} {Rev. Mod. Phys.}\
  }\textbf {\bibinfo {volume} {88}},\ \bibinfo {pages} {045006} (\bibinfo
  {year} {2016})}\BibitemShut {NoStop}%
\bibitem [{\citenamefont {Puckett}\ \emph {et~al.}(2015)\citenamefont
  {Puckett}, \citenamefont {Ni},\ and\ \citenamefont
  {Ouellette}}]{Puckett2015}%
  \BibitemOpen
  \bibfield  {author} {\bibinfo {author} {\bibfnamefont {J.~G.}\ \bibnamefont
  {Puckett}}, \bibinfo {author} {\bibfnamefont {R.}~\bibnamefont {Ni}}, \ and\
  \bibinfo {author} {\bibfnamefont {N.~T.}\ \bibnamefont {Ouellette}},\
  }\href@noop {} {\bibfield  {journal} {\bibinfo  {journal} {Phys. Rev. Lett.}\
  }\textbf {\bibinfo {volume} {114}},\ \bibinfo {pages} {258103} (\bibinfo
  {year} {2015})}\BibitemShut {NoStop}%
\bibitem [{\citenamefont {Sinhuber}\ and\ \citenamefont
  {Ouellette}(2017)}]{Sinhuber2017}%
  \BibitemOpen
  \bibfield  {author} {\bibinfo {author} {\bibfnamefont {M.}~\bibnamefont
  {Sinhuber}}\ and\ \bibinfo {author} {\bibfnamefont {N.~T.}\ \bibnamefont
  {Ouellette}},\ }\href@noop {} {\bibfield  {journal} {\bibinfo  {journal}
  {Phys. Rev. Lett.}\ }\textbf {\bibinfo {volume} {119}},\ \bibinfo {pages}
  {178003} (\bibinfo {year} {2017})}\BibitemShut {NoStop}%
\bibitem [{\citenamefont {Giavazzi}\ \emph {et~al.}(2018)\citenamefont
  {Giavazzi}, \citenamefont {Paoluzzi}, \citenamefont {Macchi}, \citenamefont
  {Bi}, \citenamefont {Scita}, \citenamefont {Manning}, \citenamefont
  {Cerbino},\ and\ \citenamefont {Marchetti}}]{Giavazzi2018}%
  \BibitemOpen
  \bibfield  {author} {\bibinfo {author} {\bibfnamefont {F.}~\bibnamefont
  {Giavazzi}}, \bibinfo {author} {\bibfnamefont {M.}~\bibnamefont {Paoluzzi}},
  \bibinfo {author} {\bibfnamefont {M.}~\bibnamefont {Macchi}}, \bibinfo
  {author} {\bibfnamefont {D.}~\bibnamefont {Bi}}, \bibinfo {author}
  {\bibfnamefont {G.}~\bibnamefont {Scita}}, \bibinfo {author} {\bibfnamefont
  {M.~L.}\ \bibnamefont {Manning}}, \bibinfo {author} {\bibfnamefont
  {R.}~\bibnamefont {Cerbino}}, \ and\ \bibinfo {author} {\bibfnamefont
  {M.~C.}\ \bibnamefont {Marchetti}},\ }\href@noop {} {\bibfield  {journal}
  {\bibinfo  {journal} {Soft Matter}\ }\textbf {\bibinfo {volume} {14}},\
  \bibinfo {pages} {3471} (\bibinfo {year} {2018})}\BibitemShut {NoStop}%
\bibitem [{\citenamefont {Cates}(2012)}]{Cates2012}%
  \BibitemOpen
  \bibfield  {author} {\bibinfo {author} {\bibfnamefont {M.~E.}\ \bibnamefont
  {Cates}},\ }\href@noop {} {\bibfield  {journal} {\bibinfo  {journal} {Rep.
  Prog. Phys.}\ }\textbf {\bibinfo {volume} {75}},\ \bibinfo {pages} {042601}
  (\bibinfo {year} {2012})}\BibitemShut {NoStop}%
\bibitem [{\citenamefont {Sanchez}\ \emph {et~al.}(2012)\citenamefont
  {Sanchez}, \citenamefont {Chen}, \citenamefont {DeCamp}, \citenamefont
  {Heymann},\ and\ \citenamefont {Dogic}}]{Sanchez2012}%
  \BibitemOpen
  \bibfield  {author} {\bibinfo {author} {\bibfnamefont {T.}~\bibnamefont
  {Sanchez}}, \bibinfo {author} {\bibfnamefont {D.~T.}\ \bibnamefont {Chen}},
  \bibinfo {author} {\bibfnamefont {S.~J.}\ \bibnamefont {DeCamp}}, \bibinfo
  {author} {\bibfnamefont {M.}~\bibnamefont {Heymann}}, \ and\ \bibinfo
  {author} {\bibfnamefont {Z.}~\bibnamefont {Dogic}},\ }\href@noop {}
  {\bibfield  {journal} {\bibinfo  {journal} {Nature}\ }\textbf {\bibinfo
  {volume} {491}},\ \bibinfo {pages} {431} (\bibinfo {year}
  {2012})}\BibitemShut {NoStop}%
\bibitem [{\citenamefont {Keber}\ \emph {et~al.}(2014)\citenamefont {Keber},
  \citenamefont {Loiseau}, \citenamefont {Sanchez}, \citenamefont {DeCamp},
  \citenamefont {Giomi}, \citenamefont {Bowick}, \citenamefont {Marchetti},
  \citenamefont {Dogic},\ and\ \citenamefont {Bausch}}]{Keber2014}%
  \BibitemOpen
  \bibfield  {author} {\bibinfo {author} {\bibfnamefont {F.~C.}\ \bibnamefont
  {Keber}}, \bibinfo {author} {\bibfnamefont {E.}~\bibnamefont {Loiseau}},
  \bibinfo {author} {\bibfnamefont {T.}~\bibnamefont {Sanchez}}, \bibinfo
  {author} {\bibfnamefont {S.~J.}\ \bibnamefont {DeCamp}}, \bibinfo {author}
  {\bibfnamefont {L.}~\bibnamefont {Giomi}}, \bibinfo {author} {\bibfnamefont
  {M.~J.}\ \bibnamefont {Bowick}}, \bibinfo {author} {\bibfnamefont {M.~C.}\
  \bibnamefont {Marchetti}}, \bibinfo {author} {\bibfnamefont {Z.}~\bibnamefont
  {Dogic}}, \ and\ \bibinfo {author} {\bibfnamefont {A.~R.}\ \bibnamefont
  {Bausch}},\ }\href@noop {} {\bibfield  {journal} {\bibinfo  {journal}
  {Science}\ }\textbf {\bibinfo {volume} {345}},\ \bibinfo {pages} {1135}
  (\bibinfo {year} {2014})}\BibitemShut {NoStop}%
\bibitem [{\citenamefont {Jee}\ \emph {et~al.}(2018)\citenamefont {Jee},
  \citenamefont {Dutta}, \citenamefont {Cho}, \citenamefont {Tlusty},\ and\
  \citenamefont {Granick}}]{Jee2018}%
  \BibitemOpen
  \bibfield  {author} {\bibinfo {author} {\bibfnamefont {A.-Y.}\ \bibnamefont
  {Jee}}, \bibinfo {author} {\bibfnamefont {S.}~\bibnamefont {Dutta}}, \bibinfo
  {author} {\bibfnamefont {Y.-K.}\ \bibnamefont {Cho}}, \bibinfo {author}
  {\bibfnamefont {T.}~\bibnamefont {Tlusty}}, \ and\ \bibinfo {author}
  {\bibfnamefont {S.}~\bibnamefont {Granick}},\ }\href@noop {} {\bibfield
  {journal} {\bibinfo  {journal} {Proc. Natl. Acad. Sci. USA}\ }\textbf
  {\bibinfo {volume} {115}},\ \bibinfo {pages} {14} (\bibinfo {year}
  {2018})}\BibitemShut {NoStop}%
\bibitem [{\citenamefont {Costerton}\ \emph {et~al.}(1999)\citenamefont
  {Costerton}, \citenamefont {Stewart},\ and\ \citenamefont
  {Greenberg}}]{Costerton1999}%
  \BibitemOpen
  \bibfield  {author} {\bibinfo {author} {\bibfnamefont {J.~W.}\ \bibnamefont
  {Costerton}}, \bibinfo {author} {\bibfnamefont {P.~S.}\ \bibnamefont
  {Stewart}}, \ and\ \bibinfo {author} {\bibfnamefont {E.~P.}\ \bibnamefont
  {Greenberg}},\ }\href@noop {} {\bibfield  {journal} {\bibinfo  {journal}
  {Science}\ }\textbf {\bibinfo {volume} {284}},\ \bibinfo {pages} {1318}
  (\bibinfo {year} {1999})}\BibitemShut {NoStop}%
\bibitem [{\citenamefont {Drescher}\ \emph {et~al.}(2011)\citenamefont
  {Drescher}, \citenamefont {Dunkel}, \citenamefont {Cisneros}, \citenamefont
  {Ganguly},\ and\ \citenamefont {Goldstein}}]{Drescher2011}%
  \BibitemOpen
  \bibfield  {author} {\bibinfo {author} {\bibfnamefont {K.}~\bibnamefont
  {Drescher}}, \bibinfo {author} {\bibfnamefont {J.}~\bibnamefont {Dunkel}},
  \bibinfo {author} {\bibfnamefont {L.~H.}\ \bibnamefont {Cisneros}}, \bibinfo
  {author} {\bibfnamefont {S.}~\bibnamefont {Ganguly}}, \ and\ \bibinfo
  {author} {\bibfnamefont {R.~E.}\ \bibnamefont {Goldstein}},\ }\href@noop {}
  {\bibfield  {journal} {\bibinfo  {journal} {Proc. Natl. Acad. Sci. USA}\
  }\textbf {\bibinfo {volume} {108}},\ \bibinfo {pages} {10940} (\bibinfo
  {year} {2011})}\BibitemShut {NoStop}%
\bibitem [{\citenamefont {Heddergott}\ \emph {et~al.}(2012)\citenamefont
  {Heddergott}, \citenamefont {Kr{\"u}ger}, \citenamefont {Babu}, \citenamefont
  {Wei}, \citenamefont {Stellamanns}, \citenamefont {Uppaluri}, \citenamefont
  {Pfohl}, \citenamefont {Stark},\ and\ \citenamefont
  {Engstler}}]{Heddergott2012}%
  \BibitemOpen
  \bibfield  {author} {\bibinfo {author} {\bibfnamefont {N.}~\bibnamefont
  {Heddergott}}, \bibinfo {author} {\bibfnamefont {T.}~\bibnamefont
  {Kr{\"u}ger}}, \bibinfo {author} {\bibfnamefont {S.~B.}\ \bibnamefont
  {Babu}}, \bibinfo {author} {\bibfnamefont {A.}~\bibnamefont {Wei}}, \bibinfo
  {author} {\bibfnamefont {E.}~\bibnamefont {Stellamanns}}, \bibinfo {author}
  {\bibfnamefont {S.}~\bibnamefont {Uppaluri}}, \bibinfo {author}
  {\bibfnamefont {T.}~\bibnamefont {Pfohl}}, \bibinfo {author} {\bibfnamefont
  {H.}~\bibnamefont {Stark}}, \ and\ \bibinfo {author} {\bibfnamefont
  {M.}~\bibnamefont {Engstler}},\ }\href@noop {} {\bibfield  {journal}
  {\bibinfo  {journal} {PLoS Pathog.}\ }\textbf {\bibinfo {volume} {8}},\
  \bibinfo {pages} {e1003023} (\bibinfo {year} {2012})}\BibitemShut {NoStop}%
\bibitem [{\citenamefont {Eisenbach}\ and\ \citenamefont
  {Giojalas}(2006)}]{Eisenbach2006}%
  \BibitemOpen
  \bibfield  {author} {\bibinfo {author} {\bibfnamefont {M.}~\bibnamefont
  {Eisenbach}}\ and\ \bibinfo {author} {\bibfnamefont {L.~C.}\ \bibnamefont
  {Giojalas}},\ }\href@noop {} {\bibfield  {journal} {\bibinfo  {journal} {Nat.
  Rev. Mol. Cell Biol.}\ }\textbf {\bibinfo {volume} {7}},\ \bibinfo {pages}
  {276} (\bibinfo {year} {2006})}\BibitemShut {NoStop}%
\bibitem [{\citenamefont {Guidobaldi}\ \emph {et~al.}(2014)\citenamefont
  {Guidobaldi}, \citenamefont {Jeyaram}, \citenamefont {Berdakin},
  \citenamefont {Moshchalkov}, \citenamefont {Condat}, \citenamefont {Marconi},
  \citenamefont {Giojalas},\ and\ \citenamefont {Silhanek}}]{Guidobaldi2014}%
  \BibitemOpen
  \bibfield  {author} {\bibinfo {author} {\bibfnamefont {A.}~\bibnamefont
  {Guidobaldi}}, \bibinfo {author} {\bibfnamefont {Y.}~\bibnamefont {Jeyaram}},
  \bibinfo {author} {\bibfnamefont {I.}~\bibnamefont {Berdakin}}, \bibinfo
  {author} {\bibfnamefont {V.~V.}\ \bibnamefont {Moshchalkov}}, \bibinfo
  {author} {\bibfnamefont {C.~A.}\ \bibnamefont {Condat}}, \bibinfo {author}
  {\bibfnamefont {V.~I.}\ \bibnamefont {Marconi}}, \bibinfo {author}
  {\bibfnamefont {L.}~\bibnamefont {Giojalas}}, \ and\ \bibinfo {author}
  {\bibfnamefont {A.~V.}\ \bibnamefont {Silhanek}},\ }\href@noop {} {\bibfield
  {journal} {\bibinfo  {journal} {Phys. Rev. E}\ }\textbf {\bibinfo {volume}
  {89}},\ \bibinfo {pages} {032720} (\bibinfo {year} {2014})}\BibitemShut
  {NoStop}%
\bibitem [{\citenamefont {Denissenko}\ \emph {et~al.}(2012)\citenamefont
  {Denissenko}, \citenamefont {Kantsler}, \citenamefont {Smith},\ and\
  \citenamefont {Kirkman-Brown}}]{Denissenko2012}%
  \BibitemOpen
  \bibfield  {author} {\bibinfo {author} {\bibfnamefont {P.}~\bibnamefont
  {Denissenko}}, \bibinfo {author} {\bibfnamefont {V.}~\bibnamefont
  {Kantsler}}, \bibinfo {author} {\bibfnamefont {D.~J.}\ \bibnamefont {Smith}},
  \ and\ \bibinfo {author} {\bibfnamefont {J.}~\bibnamefont {Kirkman-Brown}},\
  }\href@noop {} {\bibfield  {journal} {\bibinfo  {journal} {Proc. Natl. Acad.
  Sci. USA}\ }\textbf {\bibinfo {volume} {109}},\ \bibinfo {pages} {8007}
  (\bibinfo {year} {2012})}\BibitemShut {NoStop}%
\bibitem [{\citenamefont {Simmchen}\ \emph {et~al.}(2016)\citenamefont
  {Simmchen}, \citenamefont {Katuri}, \citenamefont {Uspal}, \citenamefont
  {Popescu}, \citenamefont {Tasinkevych},\ and\ \citenamefont
  {S{\'a}nchez}}]{Simmchen2016}%
  \BibitemOpen
  \bibfield  {author} {\bibinfo {author} {\bibfnamefont {J.}~\bibnamefont
  {Simmchen}}, \bibinfo {author} {\bibfnamefont {J.}~\bibnamefont {Katuri}},
  \bibinfo {author} {\bibfnamefont {W.~E.}\ \bibnamefont {Uspal}}, \bibinfo
  {author} {\bibfnamefont {M.~N.}\ \bibnamefont {Popescu}}, \bibinfo {author}
  {\bibfnamefont {M.}~\bibnamefont {Tasinkevych}}, \ and\ \bibinfo {author}
  {\bibfnamefont {S.}~\bibnamefont {S{\'a}nchez}},\ }\href@noop {} {\bibfield
  {journal} {\bibinfo  {journal} {Nat. Comms.}\ }\textbf {\bibinfo {volume}
  {7}},\ \bibinfo {pages} {10598} (\bibinfo {year} {2016})}\BibitemShut
  {NoStop}%
\bibitem [{\citenamefont {Volpe}\ \emph {et~al.}(2011)\citenamefont {Volpe},
  \citenamefont {Buttinoni}, \citenamefont {Vogt}, \citenamefont
  {K{\"u}mmerer},\ and\ \citenamefont {Bechinger}}]{Volpe2011}%
  \BibitemOpen
  \bibfield  {author} {\bibinfo {author} {\bibfnamefont {G.}~\bibnamefont
  {Volpe}}, \bibinfo {author} {\bibfnamefont {I.}~\bibnamefont {Buttinoni}},
  \bibinfo {author} {\bibfnamefont {D.}~\bibnamefont {Vogt}}, \bibinfo {author}
  {\bibfnamefont {H.-J.}\ \bibnamefont {K{\"u}mmerer}}, \ and\ \bibinfo
  {author} {\bibfnamefont {C.}~\bibnamefont {Bechinger}},\ }\href@noop {}
  {\bibfield  {journal} {\bibinfo  {journal} {Soft Matter}\ }\textbf {\bibinfo
  {volume} {7}},\ \bibinfo {pages} {8810} (\bibinfo {year} {2011})}\BibitemShut
  {NoStop}%
\bibitem [{\citenamefont {Brown}\ \emph {et~al.}(2016)\citenamefont {Brown},
  \citenamefont {Vladescu}, \citenamefont {Dawson}, \citenamefont {Vissers},
  \citenamefont {Schwarz-Linek}, \citenamefont {Lintuvuori},\ and\
  \citenamefont {Poon}}]{Brown2016}%
  \BibitemOpen
  \bibfield  {author} {\bibinfo {author} {\bibfnamefont {A.~T.}\ \bibnamefont
  {Brown}}, \bibinfo {author} {\bibfnamefont {I.~D.}\ \bibnamefont {Vladescu}},
  \bibinfo {author} {\bibfnamefont {A.}~\bibnamefont {Dawson}}, \bibinfo
  {author} {\bibfnamefont {T.}~\bibnamefont {Vissers}}, \bibinfo {author}
  {\bibfnamefont {J.}~\bibnamefont {Schwarz-Linek}}, \bibinfo {author}
  {\bibfnamefont {J.~S.}\ \bibnamefont {Lintuvuori}}, \ and\ \bibinfo {author}
  {\bibfnamefont {W.~C.~K.}\ \bibnamefont {Poon}},\ }\href@noop {} {\bibfield
  {journal} {\bibinfo  {journal} {Soft Matter}\ }\textbf {\bibinfo {volume}
  {12}},\ \bibinfo {pages} {131} (\bibinfo {year} {2016})}\BibitemShut
  {NoStop}%
\bibitem [{\citenamefont {Wykes}\ \emph {et~al.}(2017)\citenamefont {Wykes},
  \citenamefont {Zhong}, \citenamefont {Tong}, \citenamefont {Adachi},
  \citenamefont {Liu}, \citenamefont {Ristroph}, \citenamefont {Ward},
  \citenamefont {Shelley},\ and\ \citenamefont {Zhang}}]{Wykes2017}%
  \BibitemOpen
  \bibfield  {author} {\bibinfo {author} {\bibfnamefont {M.~S.~D.}\
  \bibnamefont {Wykes}}, \bibinfo {author} {\bibfnamefont {X.}~\bibnamefont
  {Zhong}}, \bibinfo {author} {\bibfnamefont {J.}~\bibnamefont {Tong}},
  \bibinfo {author} {\bibfnamefont {T.}~\bibnamefont {Adachi}}, \bibinfo
  {author} {\bibfnamefont {Y.}~\bibnamefont {Liu}}, \bibinfo {author}
  {\bibfnamefont {L.}~\bibnamefont {Ristroph}}, \bibinfo {author}
  {\bibfnamefont {M.~D.}\ \bibnamefont {Ward}}, \bibinfo {author}
  {\bibfnamefont {M.~J.}\ \bibnamefont {Shelley}}, \ and\ \bibinfo {author}
  {\bibfnamefont {J.}~\bibnamefont {Zhang}},\ }\href@noop {} {\bibfield
  {journal} {\bibinfo  {journal} {Soft Matter}\ }\textbf {\bibinfo {volume}
  {13}},\ \bibinfo {pages} {4681} (\bibinfo {year} {2017})}\BibitemShut
  {NoStop}%
\bibitem [{\citenamefont {Kantsler}\ \emph {et~al.}(2013)\citenamefont
  {Kantsler}, \citenamefont {Dunkel}, \citenamefont {Polin},\ and\
  \citenamefont {Goldstein}}]{Kantsler2013ciliary}%
  \BibitemOpen
  \bibfield  {author} {\bibinfo {author} {\bibfnamefont {V.}~\bibnamefont
  {Kantsler}}, \bibinfo {author} {\bibfnamefont {J.}~\bibnamefont {Dunkel}},
  \bibinfo {author} {\bibfnamefont {M.}~\bibnamefont {Polin}}, \ and\ \bibinfo
  {author} {\bibfnamefont {R.~E.}\ \bibnamefont {Goldstein}},\ }\href@noop {}
  {\bibfield  {journal} {\bibinfo  {journal} {Proc. Natl. Acad. Sci. USA}\
  }\textbf {\bibinfo {volume} {110}},\ \bibinfo {pages} {1187} (\bibinfo {year}
  {2013})}\BibitemShut {NoStop}%
\bibitem [{\citenamefont {Contino}\ \emph {et~al.}(2015)\citenamefont
  {Contino}, \citenamefont {Lushi}, \citenamefont {Tuval}, \citenamefont
  {Kantsler},\ and\ \citenamefont {Polin}}]{Contino2015}%
  \BibitemOpen
  \bibfield  {author} {\bibinfo {author} {\bibfnamefont {M.}~\bibnamefont
  {Contino}}, \bibinfo {author} {\bibfnamefont {E.}~\bibnamefont {Lushi}},
  \bibinfo {author} {\bibfnamefont {I.}~\bibnamefont {Tuval}}, \bibinfo
  {author} {\bibfnamefont {V.}~\bibnamefont {Kantsler}}, \ and\ \bibinfo
  {author} {\bibfnamefont {M.}~\bibnamefont {Polin}},\ }\href@noop {}
  {\bibfield  {journal} {\bibinfo  {journal} {Phys. Rev. Lett.}\ }\textbf
  {\bibinfo {volume} {115}},\ \bibinfo {pages} {258102} (\bibinfo {year}
  {2015})}\BibitemShut {NoStop}%
\bibitem [{\citenamefont {Lushi}\ \emph {et~al.}(2017)\citenamefont {Lushi},
  \citenamefont {Kantsler},\ and\ \citenamefont {Goldstein}}]{Lushi2017}%
  \BibitemOpen
  \bibfield  {author} {\bibinfo {author} {\bibfnamefont {E.}~\bibnamefont
  {Lushi}}, \bibinfo {author} {\bibfnamefont {V.}~\bibnamefont {Kantsler}}, \
  and\ \bibinfo {author} {\bibfnamefont {R.~E.}\ \bibnamefont {Goldstein}},\
  }\href@noop {} {\bibfield  {journal} {\bibinfo  {journal} {Phys. Rev. E}\
  }\textbf {\bibinfo {volume} {96}},\ \bibinfo {pages} {023102} (\bibinfo
  {year} {2017})}\BibitemShut {NoStop}%
\bibitem [{\citenamefont {Spagnolie}\ \emph {et~al.}(2017)\citenamefont
  {Spagnolie}, \citenamefont {Wahl}, \citenamefont {Lukasik},\ and\
  \citenamefont {Thiffeault}}]{Spagnolie2017Billiard}%
  \BibitemOpen
  \bibfield  {author} {\bibinfo {author} {\bibfnamefont {S.~E.}\ \bibnamefont
  {Spagnolie}}, \bibinfo {author} {\bibfnamefont {C.}~\bibnamefont {Wahl}},
  \bibinfo {author} {\bibfnamefont {J.}~\bibnamefont {Lukasik}}, \ and\
  \bibinfo {author} {\bibfnamefont {J.-L.}\ \bibnamefont {Thiffeault}},\
  }\href@noop {} {\bibfield  {journal} {\bibinfo  {journal} {Physica D}\
  }\textbf {\bibinfo {volume} {341}},\ \bibinfo {pages} {33} (\bibinfo {year}
  {2017})}\BibitemShut {NoStop}%
\bibitem [{\citenamefont {Berke}\ \emph {et~al.}(2008)\citenamefont {Berke},
  \citenamefont {Turner}, \citenamefont {Berg},\ and\ \citenamefont
  {Lauga}}]{Berke2008}%
  \BibitemOpen
  \bibfield  {author} {\bibinfo {author} {\bibfnamefont {A.~P.}\ \bibnamefont
  {Berke}}, \bibinfo {author} {\bibfnamefont {L.}~\bibnamefont {Turner}},
  \bibinfo {author} {\bibfnamefont {H.~C.}\ \bibnamefont {Berg}}, \ and\
  \bibinfo {author} {\bibfnamefont {E.}~\bibnamefont {Lauga}},\ }\href@noop {}
  {\bibfield  {journal} {\bibinfo  {journal} {Phys. Rev. Lett.}\ }\textbf
  {\bibinfo {volume} {101}},\ \bibinfo {pages} {038102} (\bibinfo {year}
  {2008})}\BibitemShut {NoStop}%
\bibitem [{\citenamefont {Li}\ and\ \citenamefont
  {Tang}(2009)}]{Li2009accumulation}%
  \BibitemOpen
  \bibfield  {author} {\bibinfo {author} {\bibfnamefont {G.}~\bibnamefont
  {Li}}\ and\ \bibinfo {author} {\bibfnamefont {J.~X.}\ \bibnamefont {Tang}},\
  }\href@noop {} {\bibfield  {journal} {\bibinfo  {journal} {Phys. Rev. Lett.}\
  }\textbf {\bibinfo {volume} {103}},\ \bibinfo {pages} {078101} (\bibinfo
  {year} {2009})}\BibitemShut {NoStop}%
\bibitem [{\citenamefont {Elgeti}\ \emph {et~al.}(2010)\citenamefont {Elgeti},
  \citenamefont {Kaupp},\ and\ \citenamefont {Gompper}}]{Elgeti2010}%
  \BibitemOpen
  \bibfield  {author} {\bibinfo {author} {\bibfnamefont {J.}~\bibnamefont
  {Elgeti}}, \bibinfo {author} {\bibfnamefont {U.~B.}\ \bibnamefont {Kaupp}}, \
  and\ \bibinfo {author} {\bibfnamefont {G.}~\bibnamefont {Gompper}},\
  }\href@noop {} {\bibfield  {journal} {\bibinfo  {journal} {Biophys. J.}\
  }\textbf {\bibinfo {volume} {99}},\ \bibinfo {pages} {1018} (\bibinfo {year}
  {2010})}\BibitemShut {NoStop}%
\bibitem [{\citenamefont {Sipos}\ \emph {et~al.}(2015)\citenamefont {Sipos},
  \citenamefont {Nagy}, \citenamefont {Di~Leonardo},\ and\ \citenamefont
  {Galajda}}]{Sipos2015}%
  \BibitemOpen
  \bibfield  {author} {\bibinfo {author} {\bibfnamefont {O.}~\bibnamefont
  {Sipos}}, \bibinfo {author} {\bibfnamefont {K.}~\bibnamefont {Nagy}},
  \bibinfo {author} {\bibfnamefont {R.}~\bibnamefont {Di~Leonardo}}, \ and\
  \bibinfo {author} {\bibfnamefont {P.}~\bibnamefont {Galajda}},\ }\href@noop
  {} {\bibfield  {journal} {\bibinfo  {journal} {Phys. Rev. Lett.}\ }\textbf
  {\bibinfo {volume} {114}},\ \bibinfo {pages} {258104} (\bibinfo {year}
  {2015})}\BibitemShut {NoStop}%
\bibitem [{\citenamefont {Spagnolie}\ \emph {et~al.}(2015)\citenamefont
  {Spagnolie}, \citenamefont {Moreno-Flores}, \citenamefont {Bartolo},\ and\
  \citenamefont {Lauga}}]{Spagnolie2015}%
  \BibitemOpen
  \bibfield  {author} {\bibinfo {author} {\bibfnamefont {S.~E.}\ \bibnamefont
  {Spagnolie}}, \bibinfo {author} {\bibfnamefont {G.~R.}\ \bibnamefont
  {Moreno-Flores}}, \bibinfo {author} {\bibfnamefont {D.}~\bibnamefont
  {Bartolo}}, \ and\ \bibinfo {author} {\bibfnamefont {E.}~\bibnamefont
  {Lauga}},\ }\href@noop {} {\bibfield  {journal} {\bibinfo  {journal} {Soft
  Matter}\ }\textbf {\bibinfo {volume} {11}},\ \bibinfo {pages} {3396}
  (\bibinfo {year} {2015})}\BibitemShut {NoStop}%
\bibitem [{\citenamefont {Elgeti}\ \emph {et~al.}(2015)\citenamefont {Elgeti},
  \citenamefont {Winkler},\ and\ \citenamefont {Gompper}}]{Elgeti2015}%
  \BibitemOpen
  \bibfield  {author} {\bibinfo {author} {\bibfnamefont {J.}~\bibnamefont
  {Elgeti}}, \bibinfo {author} {\bibfnamefont {R.~G.}\ \bibnamefont {Winkler}},
  \ and\ \bibinfo {author} {\bibfnamefont {G.}~\bibnamefont {Gompper}},\
  }\href@noop {} {\bibfield  {journal} {\bibinfo  {journal} {Rep. Prog. Phys.}\
  }\textbf {\bibinfo {volume} {78}},\ \bibinfo {pages} {056601} (\bibinfo
  {year} {2015})}\BibitemShut {NoStop}%
\bibitem [{\citenamefont {Chamolly}\ \emph {et~al.}(2017)\citenamefont
  {Chamolly}, \citenamefont {Ishikawa},\ and\ \citenamefont
  {Lauga}}]{Chamolly2017}%
  \BibitemOpen
  \bibfield  {author} {\bibinfo {author} {\bibfnamefont {A.}~\bibnamefont
  {Chamolly}}, \bibinfo {author} {\bibfnamefont {T.}~\bibnamefont {Ishikawa}},
  \ and\ \bibinfo {author} {\bibfnamefont {E.}~\bibnamefont {Lauga}},\
  }\href@noop {} {\bibfield  {journal} {\bibinfo  {journal} {New J. Phys.}\
  }\textbf {\bibinfo {volume} {19}},\ \bibinfo {pages} {115001} (\bibinfo
  {year} {2017})}\BibitemShut {NoStop}%
\bibitem [{\citenamefont {Chepizhko}\ and\ \citenamefont
  {Peruani}(2013)}]{Chepizhko2013}%
  \BibitemOpen
  \bibfield  {author} {\bibinfo {author} {\bibfnamefont {O.}~\bibnamefont
  {Chepizhko}}\ and\ \bibinfo {author} {\bibfnamefont {F.}~\bibnamefont
  {Peruani}},\ }\href@noop {} {\bibfield  {journal} {\bibinfo  {journal} {Phys.
  Rev. Lett.}\ }\textbf {\bibinfo {volume} {111}},\ \bibinfo {pages} {160604}
  (\bibinfo {year} {2013})}\BibitemShut {NoStop}%
\bibitem [{\citenamefont {Bertrand}\ \emph {et~al.}(2018)\citenamefont
  {Bertrand}, \citenamefont {Zhao}, \citenamefont {B{\'e}nichou}, \citenamefont
  {Tailleur},\ and\ \citenamefont {Voituriez}}]{Bertrand2017}%
  \BibitemOpen
  \bibfield  {author} {\bibinfo {author} {\bibfnamefont {T.}~\bibnamefont
  {Bertrand}}, \bibinfo {author} {\bibfnamefont {Y.}~\bibnamefont {Zhao}},
  \bibinfo {author} {\bibfnamefont {O.}~\bibnamefont {B{\'e}nichou}}, \bibinfo
  {author} {\bibfnamefont {J.}~\bibnamefont {Tailleur}}, \ and\ \bibinfo
  {author} {\bibfnamefont {R.}~\bibnamefont {Voituriez}},\ }\href@noop {}
  {\bibfield  {journal} {\bibinfo  {journal} {Phys. Rev. Lett.}\ }\textbf
  {\bibinfo {volume} {120}},\ \bibinfo {pages} {198103} (\bibinfo {year}
  {2018})}\BibitemShut {NoStop}%
\bibitem [{\citenamefont {Zeitz}\ \emph {et~al.}(2017)\citenamefont {Zeitz},
  \citenamefont {Wolff},\ and\ \citenamefont {Stark}}]{Zeitz2017}%
  \BibitemOpen
  \bibfield  {author} {\bibinfo {author} {\bibfnamefont {M.}~\bibnamefont
  {Zeitz}}, \bibinfo {author} {\bibfnamefont {K.}~\bibnamefont {Wolff}}, \ and\
  \bibinfo {author} {\bibfnamefont {H.}~\bibnamefont {Stark}},\ }\href@noop {}
  {\bibfield  {journal} {\bibinfo  {journal} {Eur. Phys. J. E}\ }\textbf
  {\bibinfo {volume} {40}},\ \bibinfo {pages} {23} (\bibinfo {year}
  {2017})}\BibitemShut {NoStop}%
\bibitem [{\citenamefont {Thutupalli}\ \emph {et~al.}(2018)\citenamefont
  {Thutupalli}, \citenamefont {Geyer}, \citenamefont {Singh}, \citenamefont
  {Adhikari},\ and\ \citenamefont {Stone}}]{Thutupalli2018}%
  \BibitemOpen
  \bibfield  {author} {\bibinfo {author} {\bibfnamefont {S.}~\bibnamefont
  {Thutupalli}}, \bibinfo {author} {\bibfnamefont {D.}~\bibnamefont {Geyer}},
  \bibinfo {author} {\bibfnamefont {R.}~\bibnamefont {Singh}}, \bibinfo
  {author} {\bibfnamefont {R.}~\bibnamefont {Adhikari}}, \ and\ \bibinfo
  {author} {\bibfnamefont {H.~A.}\ \bibnamefont {Stone}},\ }\href@noop {}
  {\bibfield  {journal} {\bibinfo  {journal} {Proc. Natl. Acad. Sci. USA}\
  }\textbf {\bibinfo {volume} {115}},\ \bibinfo {pages} {5403} (\bibinfo {year}
  {2018})}\BibitemShut {NoStop}%
\bibitem [{\citenamefont {Raatz}\ \emph {et~al.}(2015)\citenamefont {Raatz},
  \citenamefont {Hintsche}, \citenamefont {Bahrs}, \citenamefont {Theves},\
  and\ \citenamefont {Beta}}]{Raatz2015}%
  \BibitemOpen
  \bibfield  {author} {\bibinfo {author} {\bibfnamefont {M.}~\bibnamefont
  {Raatz}}, \bibinfo {author} {\bibfnamefont {M.}~\bibnamefont {Hintsche}},
  \bibinfo {author} {\bibfnamefont {M.}~\bibnamefont {Bahrs}}, \bibinfo
  {author} {\bibfnamefont {M.}~\bibnamefont {Theves}}, \ and\ \bibinfo {author}
  {\bibfnamefont {C.}~\bibnamefont {Beta}},\ }\href@noop {} {\bibfield
  {journal} {\bibinfo  {journal} {Eur. Phys. J. Spec. Top.}\ }\textbf {\bibinfo
  {volume} {224}},\ \bibinfo {pages} {1185} (\bibinfo {year}
  {2015})}\BibitemShut {NoStop}%
\bibitem [{\citenamefont {Machta}\ and\ \citenamefont
  {Zwanzig}(1983)}]{Machta1983}%
  \BibitemOpen
  \bibfield  {author} {\bibinfo {author} {\bibfnamefont {J.}~\bibnamefont
  {Machta}}\ and\ \bibinfo {author} {\bibfnamefont {R.}~\bibnamefont
  {Zwanzig}},\ }\href@noop {} {\bibfield  {journal} {\bibinfo  {journal} {Phys.
  Rev. Lett.}\ }\textbf {\bibinfo {volume} {50}},\ \bibinfo {pages} {1959}
  (\bibinfo {year} {1983})}\BibitemShut {NoStop}%
\bibitem [{\citenamefont {Tailleur}\ and\ \citenamefont
  {Cates}(2008)}]{Tailleur2008}%
  \BibitemOpen
  \bibfield  {author} {\bibinfo {author} {\bibfnamefont {J.}~\bibnamefont
  {Tailleur}}\ and\ \bibinfo {author} {\bibfnamefont {M.~E.}\ \bibnamefont
  {Cates}},\ }\href@noop {} {\bibfield  {journal} {\bibinfo  {journal} {Phys.
  Rev. Lett.}\ }\textbf {\bibinfo {volume} {100}},\ \bibinfo {pages} {218103}
  (\bibinfo {year} {2008})}\BibitemShut {NoStop}%
\bibitem [{\citenamefont {Schnitzer}(1993)}]{Schnitzer1993}%
  \BibitemOpen
  \bibfield  {author} {\bibinfo {author} {\bibfnamefont {M.~J.}\ \bibnamefont
  {Schnitzer}},\ }\href@noop {} {\bibfield  {journal} {\bibinfo  {journal}
  {Phys. Rev. E}\ }\textbf {\bibinfo {volume} {48}},\ \bibinfo {pages} {2553}
  (\bibinfo {year} {1993})}\BibitemShut {NoStop}%
\bibitem [{\citenamefont {Berg}(1993)}]{Berg1993}%
  \BibitemOpen
  \bibfield  {author} {\bibinfo {author} {\bibfnamefont {H.~C.}\ \bibnamefont
  {Berg}},\ }\href@noop {} {\emph {\bibinfo {title} {Random walks in
  biology}}}\ (\bibinfo  {publisher} {Princeton University Press},\ \bibinfo
  {year} {1993})\BibitemShut {NoStop}%
\bibitem [{\citenamefont {Barton}\ and\ \citenamefont
  {Ford}(1997)}]{Barton1997}%
  \BibitemOpen
  \bibfield  {author} {\bibinfo {author} {\bibfnamefont {J.~W.}\ \bibnamefont
  {Barton}}\ and\ \bibinfo {author} {\bibfnamefont {R.~M.}\ \bibnamefont
  {Ford}},\ }\href@noop {} {\bibfield  {journal} {\bibinfo  {journal}
  {Biotechnol. Bioeng.}\ }\textbf {\bibinfo {volume} {53}},\ \bibinfo {pages}
  {487} (\bibinfo {year} {1997})}\BibitemShut {NoStop}%
\bibitem [{\citenamefont {Ford}\ and\ \citenamefont {Harvey}(2007)}]{Ford2007}%
  \BibitemOpen
  \bibfield  {author} {\bibinfo {author} {\bibfnamefont {R.~M.}\ \bibnamefont
  {Ford}}\ and\ \bibinfo {author} {\bibfnamefont {R.~W.}\ \bibnamefont
  {Harvey}},\ }\href@noop {} {\bibfield  {journal} {\bibinfo  {journal} {Adv
  Water Resour}\ }\textbf {\bibinfo {volume} {30}},\ \bibinfo {pages} {1608}
  (\bibinfo {year} {2007})}\BibitemShut {NoStop}%
\bibitem [{\citenamefont {Risken}(1984)}]{Risken1984}%
  \BibitemOpen
  \bibfield  {author} {\bibinfo {author} {\bibfnamefont {H.}~\bibnamefont
  {Risken}},\ }\href@noop {} {\emph {\bibinfo {title} {The Fokker-Planck
  equation: methods of solution and applications}}}\ (\bibinfo  {publisher}
  {Springer},\ \bibinfo {year} {1984})\BibitemShut {NoStop}%
\bibitem [{\citenamefont {Chernov}(1997)}]{Chernov1997}%
  \BibitemOpen
  \bibfield  {author} {\bibinfo {author} {\bibfnamefont {N.}~\bibnamefont
  {Chernov}},\ }\href@noop {} {\bibfield  {journal} {\bibinfo  {journal} {J.
  Stat. Phys.}\ }\textbf {\bibinfo {volume} {88}},\ \bibinfo {pages} {1}
  (\bibinfo {year} {1997})}\BibitemShut {NoStop}%
\bibitem [{Sup()}]{SuppMat}%
  \BibitemOpen
  \href@noop {} {}\bibinfo {howpublished} {\url{}}\BibitemShut {NoStop}%
\bibitem [{\citenamefont {Taktikos}\ \emph {et~al.}(2013)\citenamefont
  {Taktikos}, \citenamefont {Stark},\ and\ \citenamefont
  {Zaburdaev}}]{Taktikos2013}%
  \BibitemOpen
  \bibfield  {author} {\bibinfo {author} {\bibfnamefont {J.}~\bibnamefont
  {Taktikos}}, \bibinfo {author} {\bibfnamefont {H.}~\bibnamefont {Stark}}, \
  and\ \bibinfo {author} {\bibfnamefont {V.}~\bibnamefont {Zaburdaev}},\
  }\href@noop {} {\bibfield  {journal} {\bibinfo  {journal} {PloS One}\
  }\textbf {\bibinfo {volume} {8}},\ \bibinfo {pages} {e81936} (\bibinfo {year}
  {2013})}\BibitemShut {NoStop}%
\bibitem [{\citenamefont {Lovely}\ and\ \citenamefont
  {Dahlquist}(1975)}]{Lovely1975}%
  \BibitemOpen
  \bibfield  {author} {\bibinfo {author} {\bibfnamefont {P.~S.}\ \bibnamefont
  {Lovely}}\ and\ \bibinfo {author} {\bibfnamefont {F.~W.}\ \bibnamefont
  {Dahlquist}},\ }\href@noop {} {\bibfield  {journal} {\bibinfo  {journal} {J.
  Theor. Biol.}\ }\textbf {\bibinfo {volume} {50}},\ \bibinfo {pages} {477}
  (\bibinfo {year} {1975})}\BibitemShut {NoStop}%
\bibitem [{\citenamefont {Blanco}\ and\ \citenamefont
  {Fournier}(2003)}]{Blanco2003}%
  \BibitemOpen
  \bibfield  {author} {\bibinfo {author} {\bibfnamefont {S.}~\bibnamefont
  {Blanco}}\ and\ \bibinfo {author} {\bibfnamefont {R.}~\bibnamefont
  {Fournier}},\ }\href@noop {} {\bibfield  {journal} {\bibinfo  {journal}
  {EPL}\ }\textbf {\bibinfo {volume} {61}},\ \bibinfo {pages} {168} (\bibinfo
  {year} {2003})}\BibitemShut {NoStop}%
\end{thebibliography}
%

\end{document}


\title{Supplemental Information for ``Diffusion of active particles in a complex environment: role of surface scattering''}

\flushbottom
\maketitle

\section*{reflecting boundary condition}
\subsection*{Active Santalo's formula}
To calculate the diffusion coefficient of reflecting particles, we first considered the deterministic case ($D_R=0$). In this case, there is a formula from integral geometry, known as Santalo's formula, which provides the mean free path length of a billiard moving in an array of obstacles. If the free area available to the billiards in the lattice is given by $|Q|$, and the constraining boundary is given by $|\partial Q|$, then the mean free path is
\begin{equation}
\lambda = \frac{\pi |Q|}{|\partial Q|}.
\end{equation}

\begin{figure}
	\includegraphics[width=\textwidth]{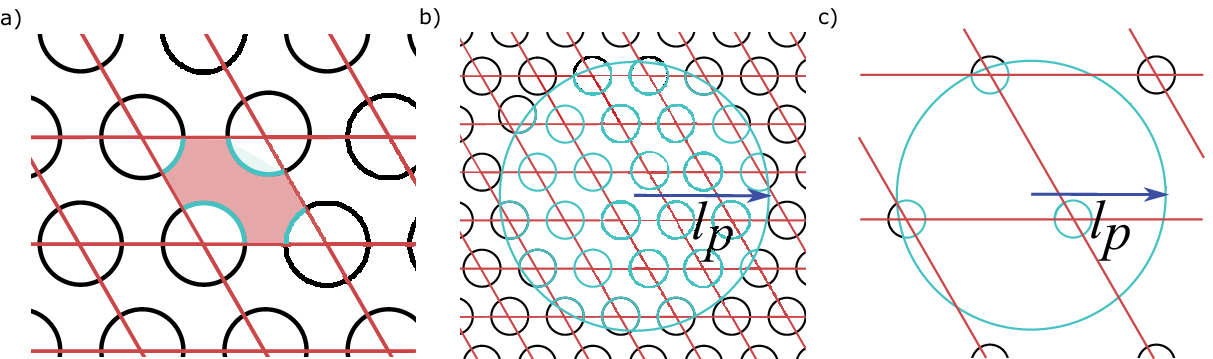}
	\caption{An illustration of the lattice geometry in the reflecting boundary condition calculation. A hexagonal lattice can be patterned by rhombi of side length equal to the lattice spacing, as shown in a). The shaded area is the area available to particles per unit cell, $A$, while the blue arcs highlight the obstacle surface per unit cell $P=2\pi R$. In b) and c), an extra circular boundary is added, with a radius of the persistence length, to account for reorientation via rotational diffusion. In c) this effect will be more marked as the obstacle density is lower.}
	\label{fig:sant}
\end{figure}

For our hexagonal lattice, we know that the ratio between a unit cell's free area, $A$ to its constraining surface, $P$ will be the same as for the whole lattice. Choosing the unit cell to be a rhombus, $A=\sqrt{3}d^2/2-\pi R^2$, and $P=2\pi R$ (see Fig.~\ref{fig:sant}a) for an illustration), and so the mean free path length then follows immediately as
\begin{equation}
\lambda = \frac{(\sqrt{3}d^2-2\pi R^2)}{4 R}.
\end{equation}
This formula should hold for active particles at high obstacle density, where the persistence length $l_p=v/D_R$ is much larger than the separation between pillars, $d-2R$. The trajectories will then almost look deterministic.

We found the diffusion coefficient $D=v \lambda /2$ only fits the simulation data when we divide the mean free path length by a factor $\pi/2$:
\begin{equation}
D = \frac{v}{2}\frac{(\sqrt{3}d^2-2\pi R^2)}{2\pi R}.
\end{equation}
where $d$ is the lattice spacing, and $R$ the radius of the circular pillars. We believe that this is due to the choice of averaging conditions made in the earlier works. As an example, if you wish to calculate the mean free path length in a confining circle, you must average over all possible chords within that circle. Depending on how you construct that average, you get different answers. So, we think that the averaging process used in the works of integral geometry are not appropriate to our case.

Although this result is good for high obstacle densities, we would like a model that works at low densities as well. At low obstacle densities, we can no longer ignore rotational diffusion. The deterministic formula must obviously diverge as $R\to 0$, as some rays will travel infinitely far before their next collision.

For active particles, the persistence length defines a length over which directional persistence is lost. This is analogous to the mean free path length in a gas. Guided by this, we include a circle with radius of the persistence length into the lattice. This acts as an extra piece of obstacle boundary, and constrains the particle according to its rotational diffusion coefficient. For high obstacle densities, this will play only a minor role; a particle is much more likely to hit an obstacle than the surrounding boundary, as illustrated in Fig.~\ref{fig:sant}b). However, when the obstacle density becomes low enough, then the circle becomes important, as in Fig.~\ref{fig:sant}c). Now, the mean free path becomes (dividing the Santalo formula by the same factor of $\pi/2$ as above):
\begin{equation}
\lambda = \frac{2NA}{NP+2\pi l_p},
\end{equation}
where $N=\pi l_p^2/(\sqrt{3}d^2/2)$ is the number of rhombus unit cells within the confining circle. Note that for $R=0$, $A=\sqrt{3}d^2/2$, and so $\lambda=l_p$ as required.

\subsection*{Diffusion with reflecting boundary condition}

\begin{figure}
	\includegraphics[width=0.5\textwidth]{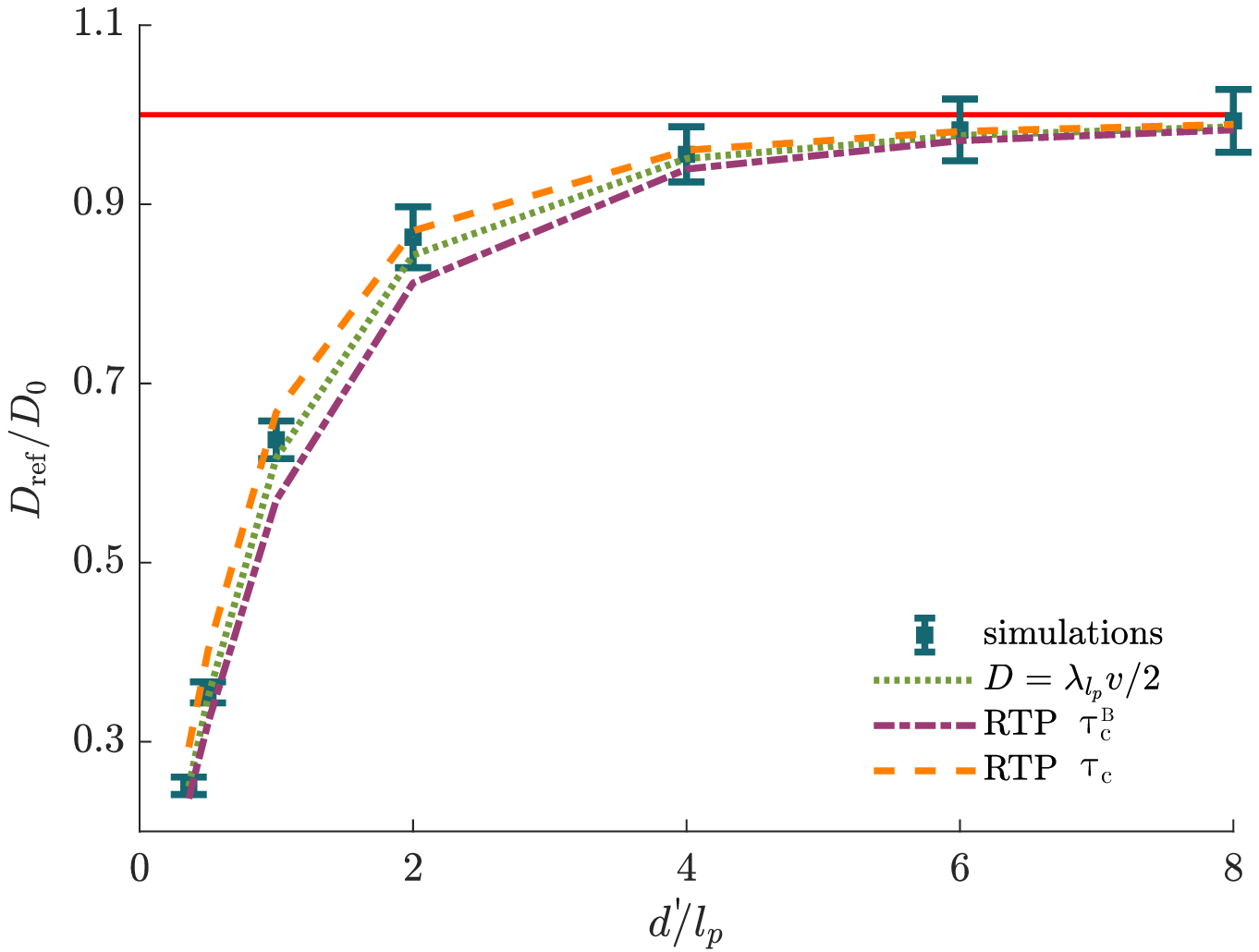}
	\caption{Diffusion coefficient of an active particle with persistence length $l_p=v/D_R$ as a function of the separation of the obstacles with a constant radius of $R=12\mu m$. The diffusion coefficient was scaled by the free diffusion coefficient in the absence of any obstacles, $D_0$.}
	\label{fig:DvsLp}
\end{figure}

Fig. 2(a) in the main text shows the diffusion coefficient with reflective boundary condition as function of the obstacle density in terms of $R/d$, where $d$ is kept constant. Fig.~\ref{fig:DvsLp} shows $D_\mathrm{ref}$ as a function of $d'/l_p$, where $d'$ is the minimum free distance between obstacles ($d'=d-2R$). For $d>>l_p$, the obstacle density is very small and $D_\mathrm{ref}$ approaches the free diffusion limit $D_0$. Both the active Santalo formula and the RTP model with Santalo free path give a good approximation of the simulation results.

\clearpage
\section*{Sliding boundary condition}
\subsection*{Collision angle distribution}
\begin{figure}[h]
\includegraphics[width=0.5\textwidth]{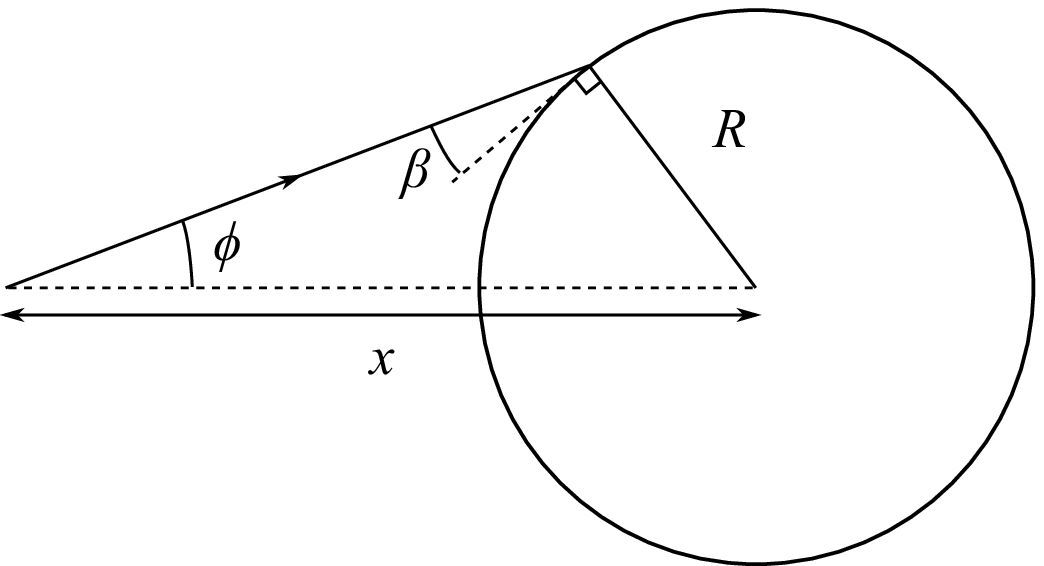}
\caption{Schematic of the set-up to determine the incoming angle distribution. A particle starts a distance $x$ from the centre of the circle, and then travels at an orientation $\phi$ relative to the horizontal, hitting the circle at an angle $\beta$ to the tangent of the circle.}
\label{fig:coll}
\end{figure}

To find the orientation function $\langle \cos\theta\rangle$, we must first determine the collision angle distribution. At low densities, the collision angle distribution can be approximated by the problem set up in Fig.~\ref{fig:coll} and described in the main text. Since there is radial symmetry, we do not have to consider a specific direction, $\mathbf{x}$, relative to the circle.
The function $\phi(\beta,x)$ defined in the main text can be easily worked out through the sine rule:
\begin{equation}
\frac{x}{\sin(\beta+\pi/2)}=\frac{R}{\sin\phi},
\end{equation}
\subsection*{Deriving the reorientation angle}
\begin{figure}
	\centering
	\includegraphics[width=0.5\textwidth]{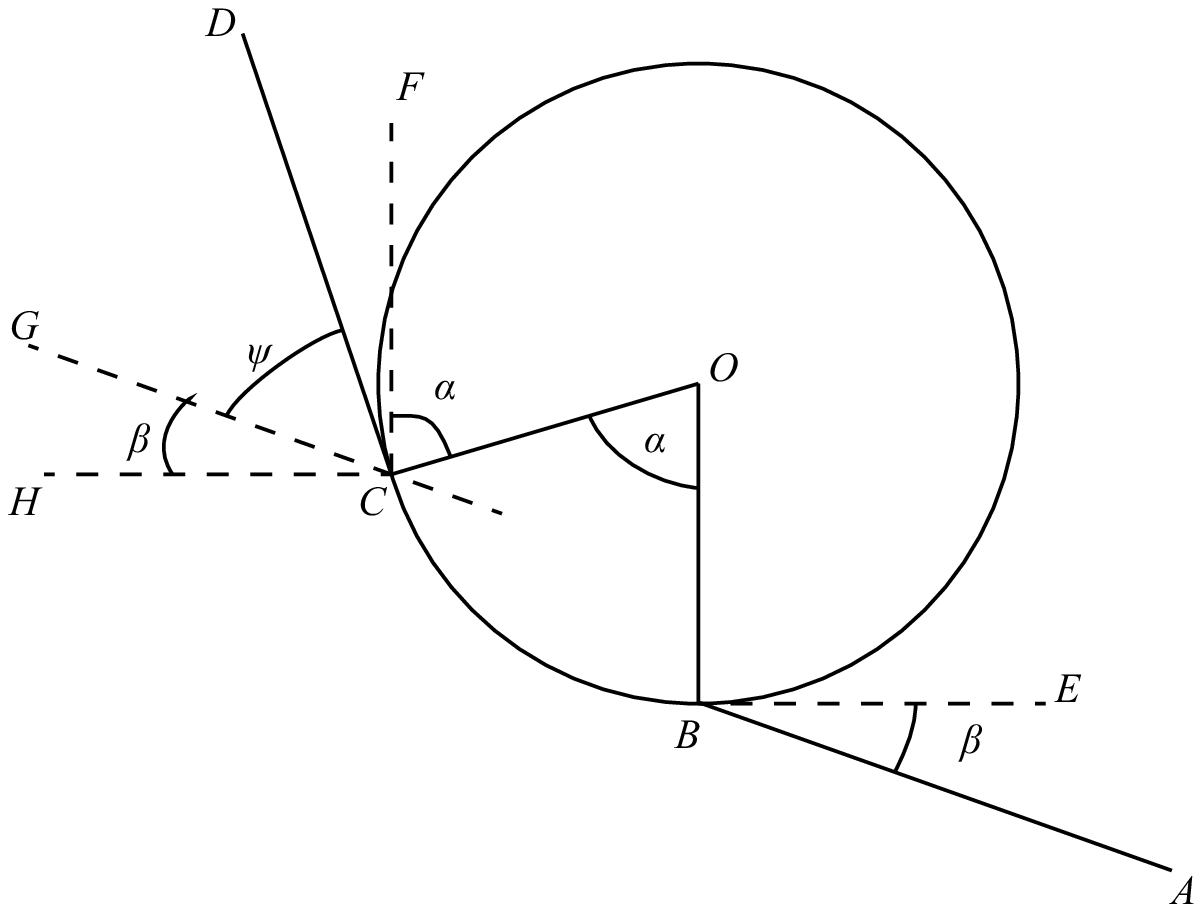}
	\caption{Schematic of the reorientation angle $ \theta$ as a function of $\alpha$ and $\beta$. The sliding particle travels along the path ABCD. Lines OB and CF are parallel, as are lines AB and CG, and lines EB and CH.}
	\label{fig:dtheta}
\end{figure}
By construction (see Fig.~\ref{fig:dtheta}), angle FCH must be a right angle. Since the particle leaves at a tangent, OCD is also a right angle. By the alternate angle theorem, OCF=$\alpha$, and so DCF=$\pi/2-\alpha$. Therefore, $ \psi=\pi/2-\beta-(\pi/2-\alpha)=\alpha-\beta$. 
\clearpage
\subsection*{Diffusion with sliding boundary condition}

\begin{figure}[h!]
	\includegraphics[width=0.5\textwidth]{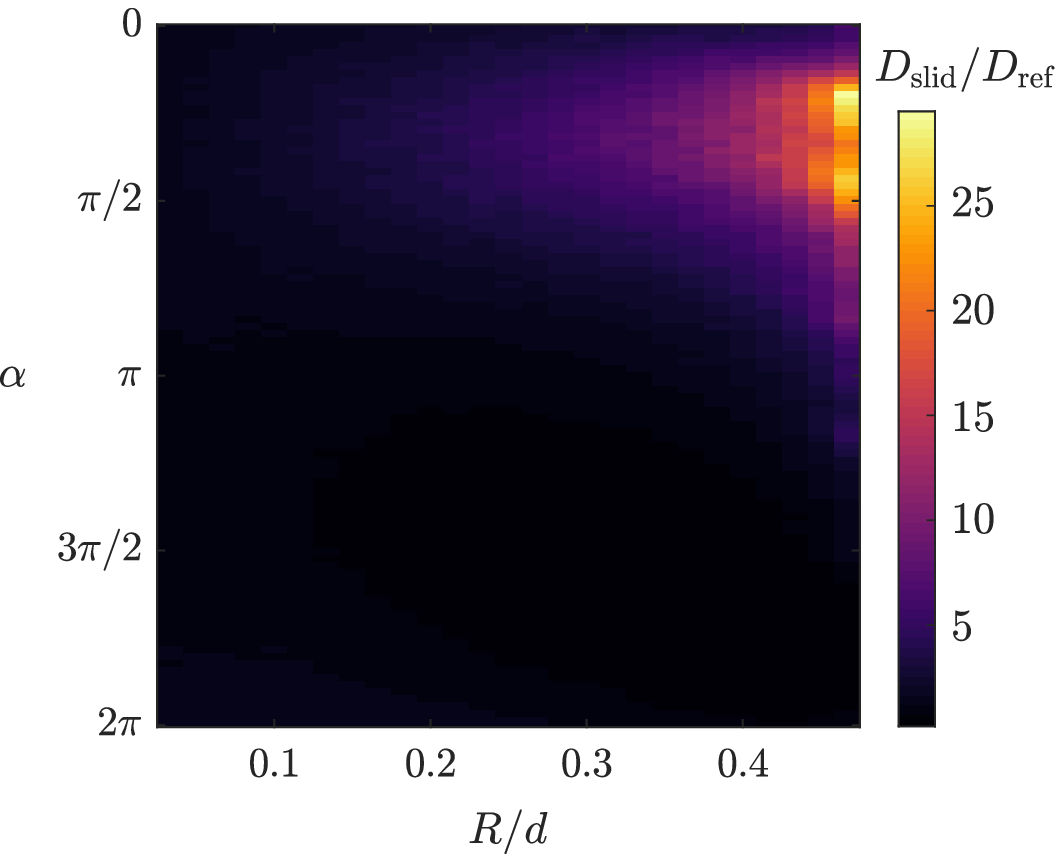}
	\caption{Diffusion in hexagonal lattice of obstacles. Same data as in figure 2b in main text but now scaled by the reflective diffusion coefficient, $D_\mathrm{ref}$, rather than the free diffusion coefficient.}
	\label{fig:DslidVsDref}
\end{figure}
\section{Geometrical guiding effect}
When the radius $R>\sqrt{3}d/4$, the lattice has a nearest neighbour horizon; any straight line drawn from the surface of a pillar must hit one of its nearest neighbours. In this case, for the sliding boundary condition, in a simplified lattice of two rows, we can find the sequence of leaving angles from pillars along a run, as detailed in the main text. The function $f(\theta_n)$ can be determined fairly easily through geometry, and takes the form
\begin{equation}
f(\theta_n)=\left\{\begin{array}{lr}\cos^{-1}\left(1+\dfrac{d}{R}\sin\left(\dfrac{\pi}{6}-\theta_n\right)\right)-\theta_n, & \theta_n < \theta^{*} \\
\theta_n-\cos^{-1}\left(1+\dfrac{d}{R}\cos\theta_n\right), & \theta_n \geq \theta^{*}
\end{array}\right.
\end{equation}
The angle $\theta^{*}=2\pi/3-\cos^{-1}(R/d)$ is a transition angle. For $\theta_n<\theta^{*}$, the particle will hit a pillar on the opposite side of the channel, and for $\theta\geq\theta^{*}$ the particle will hit on the same side of the channel.